\documentclass[epj]{svjour}

\usepackage{graphicx} 
\usepackage{subfigure} 
\usepackage{amsmath} 

\usepackage{epsfig}
\usepackage{epstopdf}

\usepackage{caption}

\usepackage{graphics}
\usepackage{graphicx}
\usepackage{grffile} 

\usepackage{amsmath}

\usepackage{color}

\usepackage{hyperref}

\usepackage[utf8]{inputenc}

\usepackage{lineno,hyperref}

\usepackage{geometry}
\newcommand{\npart}{N_{part}}
\newcommand{\ncoll}{N_{coll}}

\begin{document}

\title{A systematic study of the initial state in heavy ion collisions based on the quark participant assumption}

\author{Liang Zheng
\thanks{{e-mail:} zhengl@mails.ccnu.edu.cn}
\and Zhongbao Yin
\thanks{{e-mail:} zbyin@mail.ccnu.edu.cn}
}

\institute{
Key Laboratory of Quark and Lepton Physics (MOE) and Institute
of Particle Physics, Central China Normal University, Wuhan 430079, China
}

\date{\\}

\abstract{
We investigate the initial state geometric quantities of heavy ion collisions based on the quark participant assumption in the Glauber multiple scattering approach. A systematic comparison to the nucleon participant assumption has been presented and confronted with the charged multiplicity measurements in various collision systems. It is found that the quark participant based assumption can be important to understand the data in multiplicity production and the initial spatial eccentricity in small systems. 
}

     
\authorrunning{L. Zheng and Z.B. Yin}
\titlerunning{Initial state of heavy ion collisions with quark participant assumption}

\maketitle


\section{Introduction}
The concept of wounded nucleon for particle production has been formulated and widely applied to describe particle production from nuclear targets for many years~\cite{Bialas:1976ed,Bialas:1977en,Bialas:2012dc}. A wounded nucleon is a nucleon participant involved in the nuclear inelastic collisions. In high energy collisions on a nuclear target, it is assumed that the target nucleus is rather transparent, so that the incoming projectile constituents can independently make many successive collisions while passing through the nucleus. The inelastic collisions of two nuclei can thus be described as an incoherent composition of the collisions between individual constituents of the colliding nuclei.  

The application of the wounded nucleon concept in nuclear collisions leads to the prediction that the multiplicity production should be proportional to the number of participant nucleons in the collision. The empirical results of first accelerator measurements~\cite{Busza:1975df} bring us the observational foundation of the so called wounded nucleon model (WNM)~\cite{Bialas:1976ed}. On the other hand, the data obtained at RHIC indicate that the WNM works for the total multiplicity but fails in the description of the multiplicity density per participant pair at mid-rapidity as a function of collision centrality~\cite{Alver:2010ck}. The most common explanation for this failure involves particle production in hard processes. A two component model~\cite{Wang:2000bf} has been introduced to explain this centrality dependence, while the energy dependence of the relative contributions arising from soft and hard processes is still difficult to reconcile~\cite{Aamodt:2010cz}.

The idea of wounded quarks as a generalization of the wounded source in WNM proposed in~\cite{Bialas:1977en} can reasonably describe the mid-rapidity charged hadron density without need to incorporate the hard scattering component. In the wounded quark picture, heavy ion collisions are effectively modeled by the elementary reactions between the participating quark constituents of every nucleon from the projectile and target nucleus. This approach has been further applied to study the charged hadron density, total multiplicity~\cite{Eremin:2003qn,KumarNetrakanti:2004ym,De:2004gn,Nouicer:2006pr}, multiplicity fluctuation~\cite{Fialkowski:2010tr}, transverse energy distribution~\cite{Adler:2013aqf,Adare:2015bua} and strangeness enhancement~\cite{Behera:2012eq} in a large variety of center of mass energy and collision systems. 

In this work, we will show a systematic study when interpreting the collisions by replacing the original elementary nucleon-nucleon reactions with the quark-quark reactions from those nucleons. This paper is organized as follows: we give a systematic study on the quark participant model in sec.~\ref{sec:model_sys} and confront the model with the multiplicity production data in various collision systems in sec.~\ref{sec:data_comparison}. The applicability of this model and its impact on the understanding of initial states in heavy ion collisions will be discussed in sec.~\ref{sec:impact}. In the end, we summarize in sec.~\ref{sec:summary}.

\section{Quark participants model systematics}
\label{sec:model_sys}

\begin{figure}[htbp]
\centering
\includegraphics[width=0.55\textwidth]{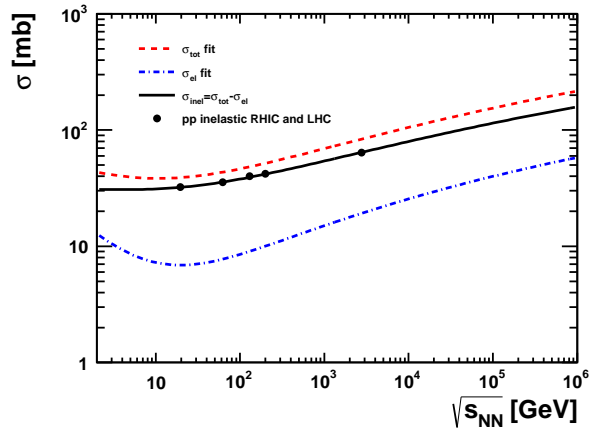}
\caption{(Color online) Interpolation of the inelastic cross section (solid line) based on the measured pp data by subtracting the elastic contribution (dash-dotted line) from total cross section (dashed line).  Total and elastic cross section parameterizations are from PDG data~\cite{Beringer:1900zz}, which have also been used in~\cite{Abelev:2013qoq}. The interpolated inelastic cross section agrees reasonably with the inelastic cross section measurements from RHIC~\cite{Abelev:2008ab} and LHC~\cite{Abelev:2013qoq}.}
\label{fig:inel_cross_section}
\end{figure}

\begin{figure}[htbp]
\centering
\includegraphics[width=0.55\textwidth]{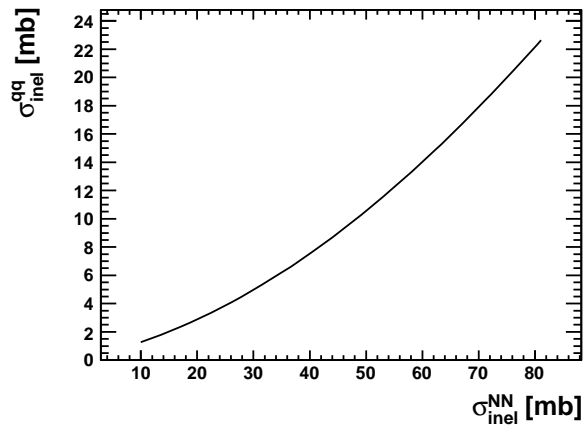}
\caption{Inelastic quark-quark cross section $\sigma^{qq}_{inel}$ adjuested to different inelastic nucleon-nucleon cross section $\sigma^{NN}_{inel}$.}\label{fig:sigQQ_vs_sigNN}
\end{figure}

A standard Monte Carlo Glauber calculation~\cite{Miller:2007ri,Loizides:2014vua} is performed to obtain the mean number of the nucleon and quark participants inside a nucleus. We assemble the nucleus by sampling the initial positions of the nucleons inside the nucleus based on a three-parameter Fermi (3pF) distribution
\begin{equation}
\rho^A (R) = \rho_0\frac{1+w(R/R_{0})^2}{1+\exp[(R-R_{0})/d]},
\label{eqn:wood_saxon}
\end{equation}
in which $R$ gives the radial position of a nucleon, $\rho_0$ represents the nucleon density in a nucleus, $R_{0}$ is the nuclear radius, $d$ is the skin depth and $w$ shows the deviations from a spherical shape.
Following the procedure suggested in~\cite{Adler:2013aqf,Adare:2015bua}, three quarks are then sampled with an empirical functional form in the radial direction:
\begin{eqnarray}
f(r)= & r^{2}e^{-4.27r}(1.21466-1.888r+2.03r^{2}) \nonumber \\
      & (1+\frac{1}{r}-\frac{0.03}{r^{2}})(1+0.15r), 
\label{eqn:empirical_form}
\end{eqnarray}
where $r$ shows the distance of a quark to its parent nucleon center obtained from Eq.~\ref{eqn:wood_saxon}. When the positions of three quarks are determined, we shift the center of mass of the three quark system back to the coordinate of the nucleon.
The empirical function chosen above is supposed to correct the shift effect to guarantee the quark radial distribution reproduces the Fourier transform of the form factor of proton~\cite{Hofstadter:1958}:
\begin{equation}
\rho^{proton} (r)= \rho^{proton}_0 \times e^{-ar},
\label{eqn:proton_form}
\end{equation}
where $a=\sqrt{12}/r_{m}=4.27 \mathrm{fm}^{-1}$ and the RMS charge radius of proton $r_m=0.81 \mathrm{fm}$. Assigned with a relative displacement according to the impact parameter, one elementary reaction happens between the participants when their distance $d$ in the transverse plane satisfies the criteria:
\begin{equation}
d < \sqrt{\frac{\sigma_{inel}}{\pi}}.
\end{equation}
$\sigma_{inel}$ represents the energy dependent elementary cross section between the two colliding constituents. The nucleon-nucleon inelastic cross section $\sigma^{NN}_{inel}$ can be estimated by interpolation of pp data at a wide range of center-of-mass energy and subtracting the elastic cross section from the total cross section, as shown in Fig.~\ref{fig:inel_cross_section}.
The inelastic quark-quark cross section $\sigma^{qq}_{inel}$ is obtained by adjusting for the nucleon-nucleon collisions until the known  nucleon-nucleon cross section at certain energy scale is reproduced. The correspondence of $\sigma^{qq}_{inel}$ to $\sigma^{NN}_{inel}$ can be found in Fig.~\ref{fig:sigQQ_vs_sigNN}. As the elementary cross section $\sigma^{NN}_{inel}$ in our concerned energy range varies in 30-70 mb, the relevant quark-quark cross section in our study ranges from 5 to 18 mb.

\begin{figure}[htbp]
\centering
\includegraphics[width=0.55\textwidth]{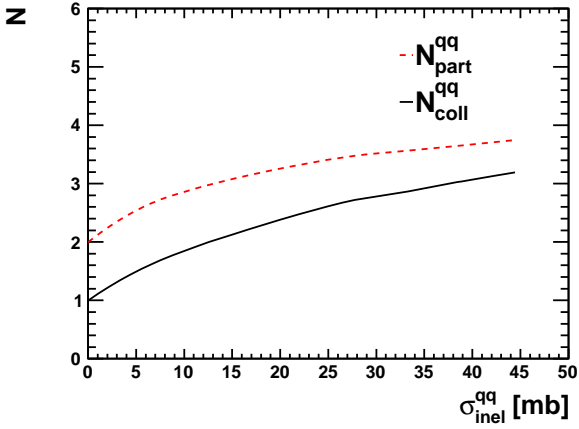}
\caption{(Color online) $\npart$ (dashed line) and $\ncoll$ (solid line) for quark participants in pp collisions based on different quark-quark cross section.}\label{fig:npart_ncoll}
\end{figure}

With the convolution of quark-quark cross section and nucleon-nucleon cross section defined above, one can estimate the mean number of quark participants and collisions in the pp collisions dependent on the quark-quark inelastic cross section, which can be found in Fig.~\ref{fig:npart_ncoll}.
It is observed that for the case of pp collisions, the quark participant number over the number of binary collision is around two at very small $\sigma^{qq}_{inel}$. As energy increases, the growth of quark-quark cross section makes it possible to have more than one collision on one projectile, thus the ratio of $N_{part}/N_{coll}$ becomes less than two for large $\sigma^{qq}_{inel}$.

\begin{figure}[htbp]
\centering
\includegraphics[width=0.55\textwidth]{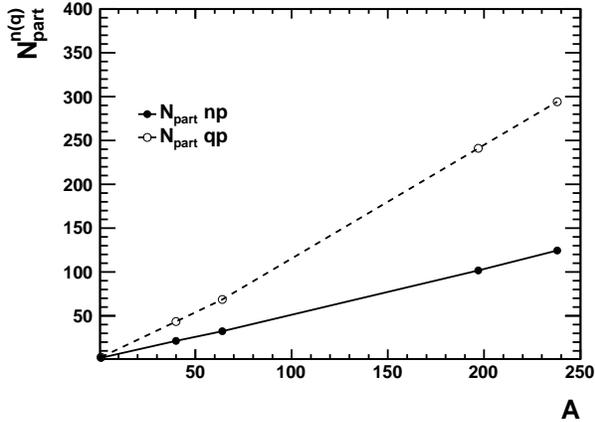}
\caption{$\npart$ for pp, CaCa, CuCu, AuAu, UU collisions at $\sqrt{s_{NN}}=200$ GeV based on nucleon (full marker) and quark (empty marker) participants assumption. qp represents quark participant while np represents nucleon participant.}\label{fig:npart_sys_size}
\end{figure}

Fig.~\ref{fig:npart_sys_size} shows the dependence of $\npart$ on the systems size for the minimum bias events in both nucleon and quark participant assumptions at a fixed energy of $\sqrt{s_{NN}}=200$ GeV. It is expected that $\npart$ should scale with the volume of interaction region, therefore $\npart$ is supposed to be proportional to $A$, as is shown in this figure.

\begin{figure}[htbp]
\centering
\includegraphics[width=0.55\textwidth]{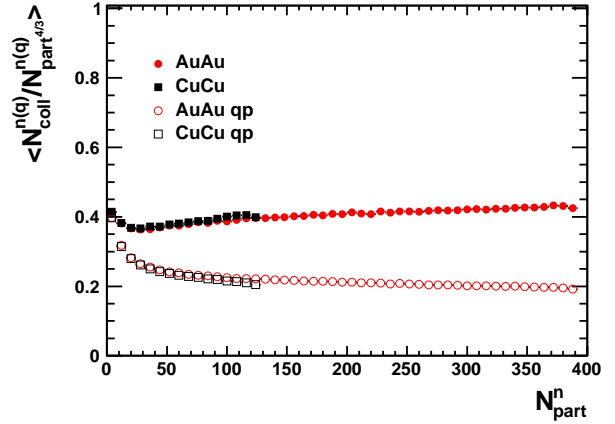}
\caption{(Color online) Geometric scaling behavior between $\npart$ and $\ncoll$ for quark participants in CuCu (square) and AuAu (circle) collisions at $\sqrt{s_{NN}}=200$ GeV based on nucleon (solid marker) and quark (open marker) participants assumption.}\label{fig:geo_scale}
\end{figure}

It is also noticed that the number of binary collisions experienced per participant scale with the length of the interaction volume along the projectile moving direction $l_z \propto N^{1/3}_{part}$ so that we expect the geometric scaling $N_{coll} \propto N^{4/3}_{part}$, which has been shown in Fig.~\ref{fig:geo_scale} for both nucleon and quark participants.

\section{Relating quark participant assumptions to multiplicity measurements}
\label{sec:data_comparison}
It is believed that charged particle multiplicity densities near mid-rapidity in high energy nuclear collisions scale with the number of participant pairs involved in the collisions. We have carried out the study for the dependence of the charged multiplicity on $N_{part}$ in nucleon-nucleon collision framework and quark-quark collision framework.

\begin{figure}[htbp]
\centering
\includegraphics[width=0.55\textwidth]{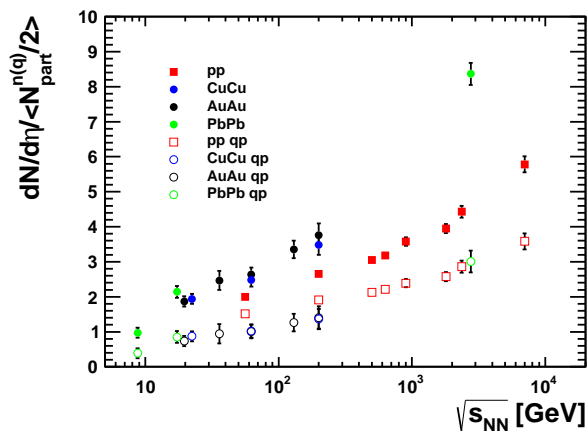}
\caption{(Color online) Energy dependence of the charged particle production with nucleon-nucleon collision pairs and quark-quark collision pairs. Full dots indicate that the charged particle density scaled by the number of nucleon participants, while the open circles are that scaled by the quark constituent participant number in corresponding nucleus-nucleus collisions. Red filled squares represent the NSD pp or $\bar{\mathrm{p}}$p data from ~\cite{Alver:2010ck,Khachatryan:2010us,Abe:1989td,Ansorge:1988kn} and quark participant number normalized $pp$ result is shown in open squares. qp represents quark participant. Compiled data for nucleus-nucleus collisions come from Ref.~\cite{Alver:2010ck,Aamodt:2010cz,Abreu:2002fw}. }
\label{fig:dNdEta_energy}
\end{figure}
It is shown in Fig.~\ref{fig:dNdEta_energy} that, the charged particle densities for heavy ions and pp or $\mathrm{p}\bar{\mathrm{p}}$ are splitting when scaled with the number of nucleon participants, while that scaled by the number of quark participants is following the same trend. It implies that the elementary reactions between participants can be better described by the quark constituents. 
\begin{figure*}
\centering
\subfigure[]{
\includegraphics[width=0.45\textwidth]{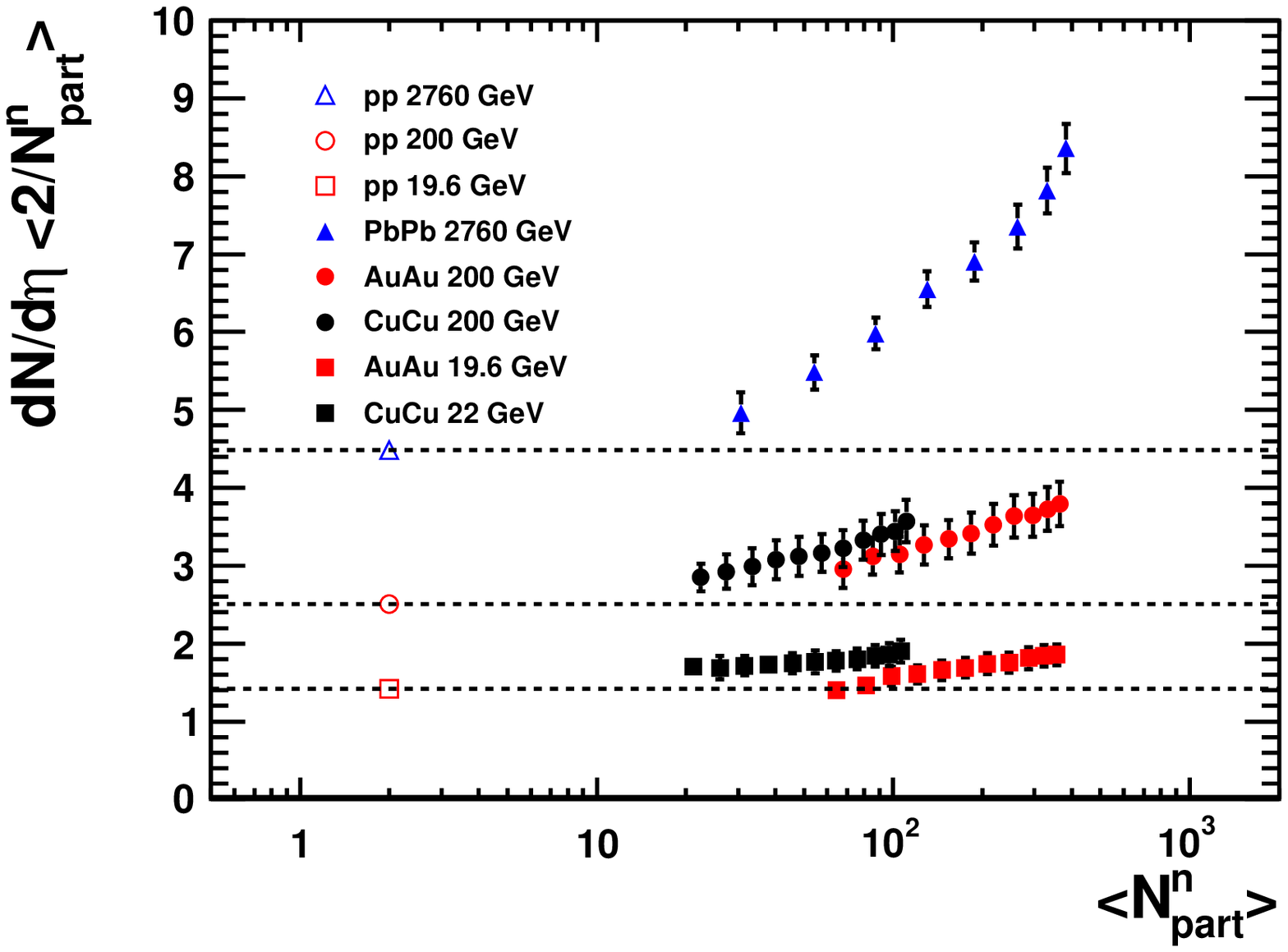}
\label{fig:dNdEta_centrality_normal}
}
\subfigure[]{
\includegraphics[width=0.45\textwidth]{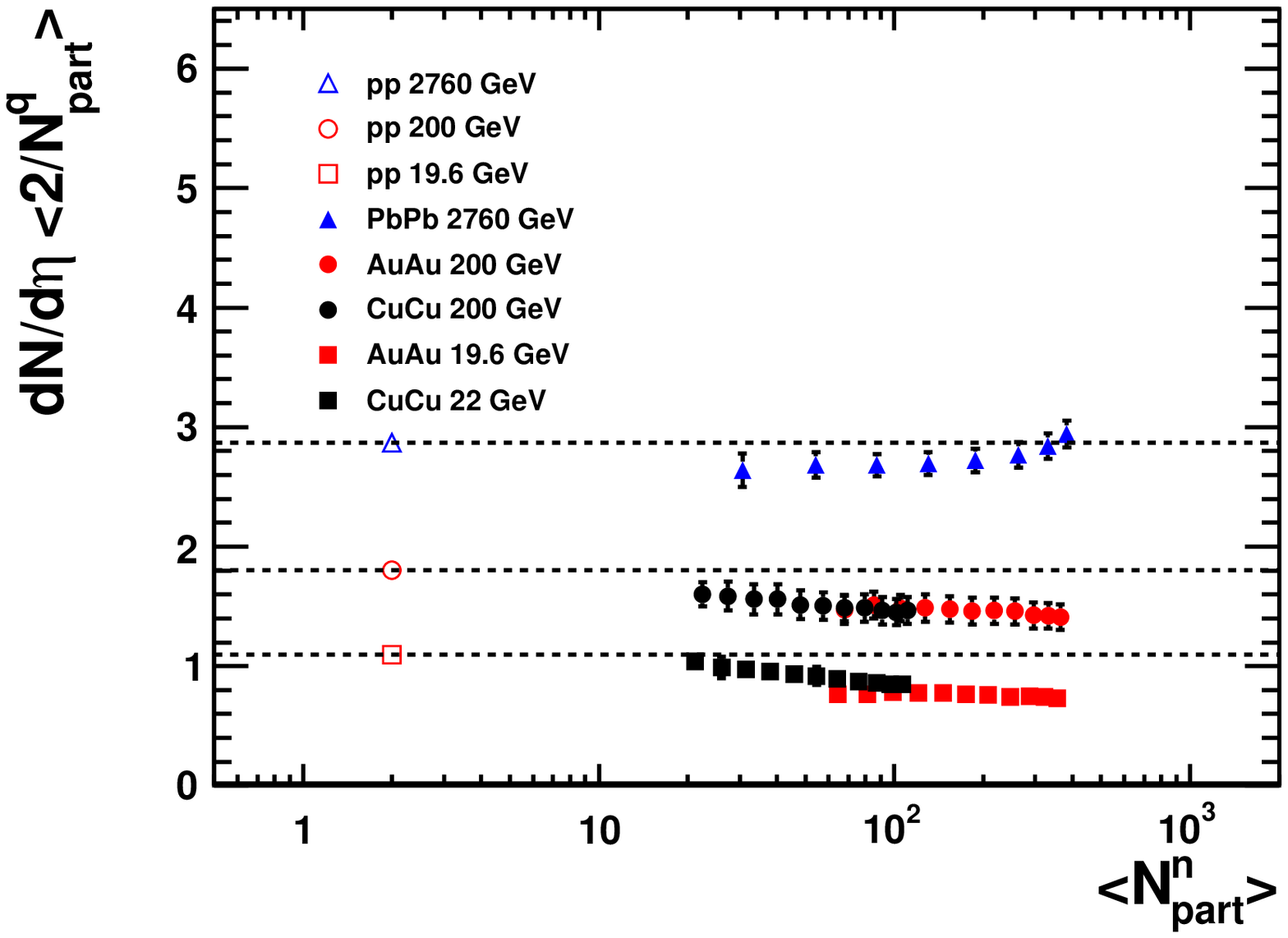}
\label{fig:dNdEta_centrality_ncq}
}
\caption{(Color online) Mid-rapidity charged particle density varying with centrality in different collision systems scaled by the number of nucleon participants. Left panel shows the result for the nucleon participant case while quark participant result is shown in the right panel. pp data points are extrapolated from the parameterizations in Ref.~\cite{Chatrchyan:2011pb}. The heavy ion results on $dN/d\eta$ for PbPb, AuAu and CuCu come from the Ref.~\cite{Aamodt:2010cz,Alver:2010ck}. }
\label{fig:dNdEta_centrality}
\end{figure*}
As shown in Fig.~\ref{fig:dNdEta_centrality_normal}, the charged particle density per participant pair grows from peripheral collisions to central collisions especially taking the pp extrapolated points into consideration, when normalized by the nucleon participants. The significant growth in charged density clearly deviates from wounded nucleon model assumption. Inspired by the two component model~\cite{Wang:2000bf}, the centrality dependence has often been interpreted by a combination of particle production from soft and hard processes. On the other hand, the charged density is showing the same shape versus $N_{part}$ as varied in a wide range of energy scale ranging from 7.7 GeV to 2760 GeV when the hard scattering cross section changes more than a factor of twenty~\cite{Tannenbaum:2015zha}. This observation leads to an argument that the hard scattering component can hardly be responsible for the nuclear geometrical effect. Whereas one observes a slightly varying density when scaled by the quark participant number in Fig.~\ref{fig:dNdEta_centrality_ncq}, except the low energy data points. This suggests the charged particle density in high energy collisions is approximately proportional to the number of quark participants and reconciles the explanation with hard scattering component. 

\begin{figure*}
\centering
\subfigure[]{
\includegraphics[width=0.45\textwidth]{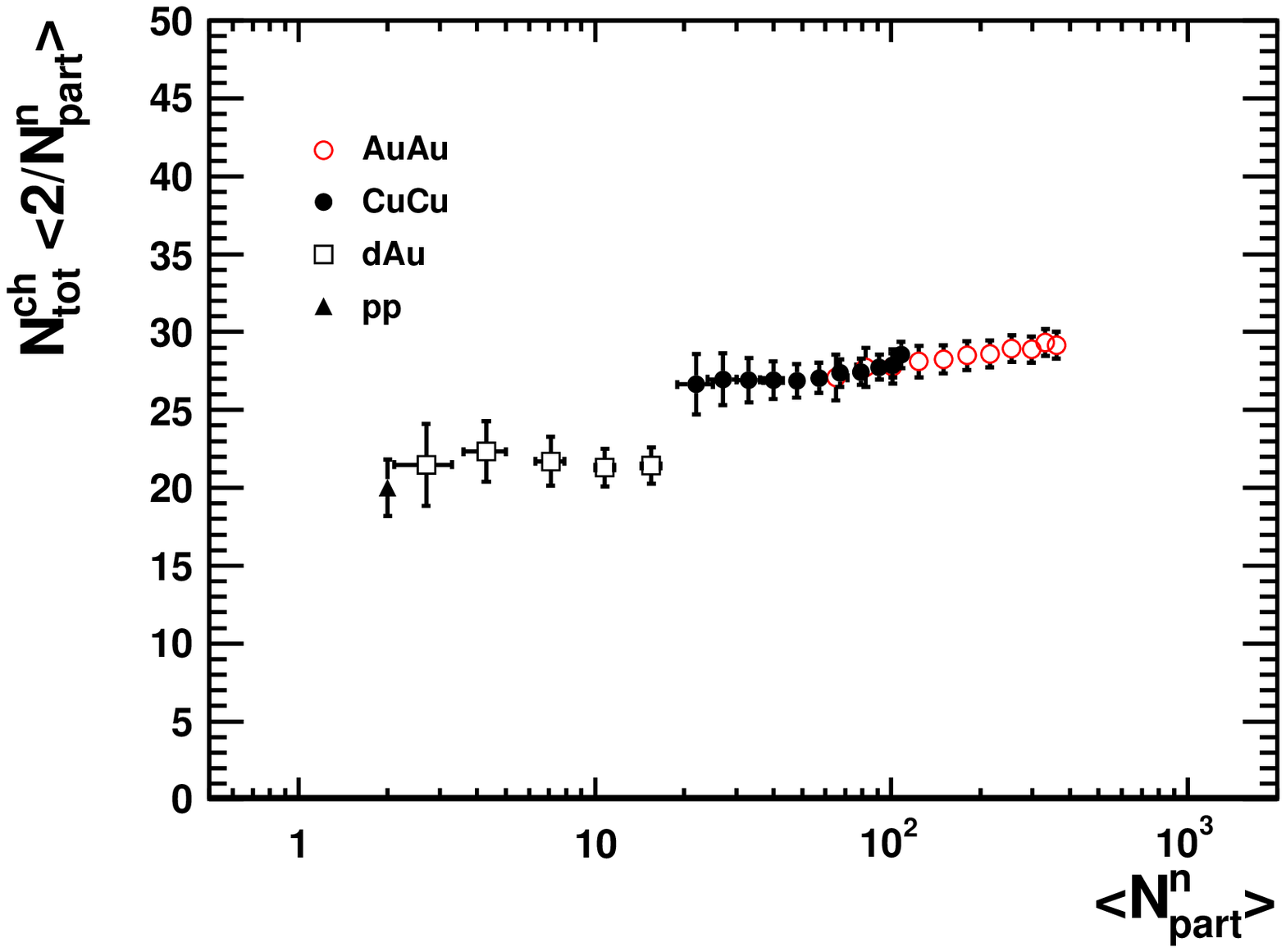}
\label{fig:Nch_tot_centrality_normal}
}
\subfigure[]{
\includegraphics[width=0.45\textwidth]{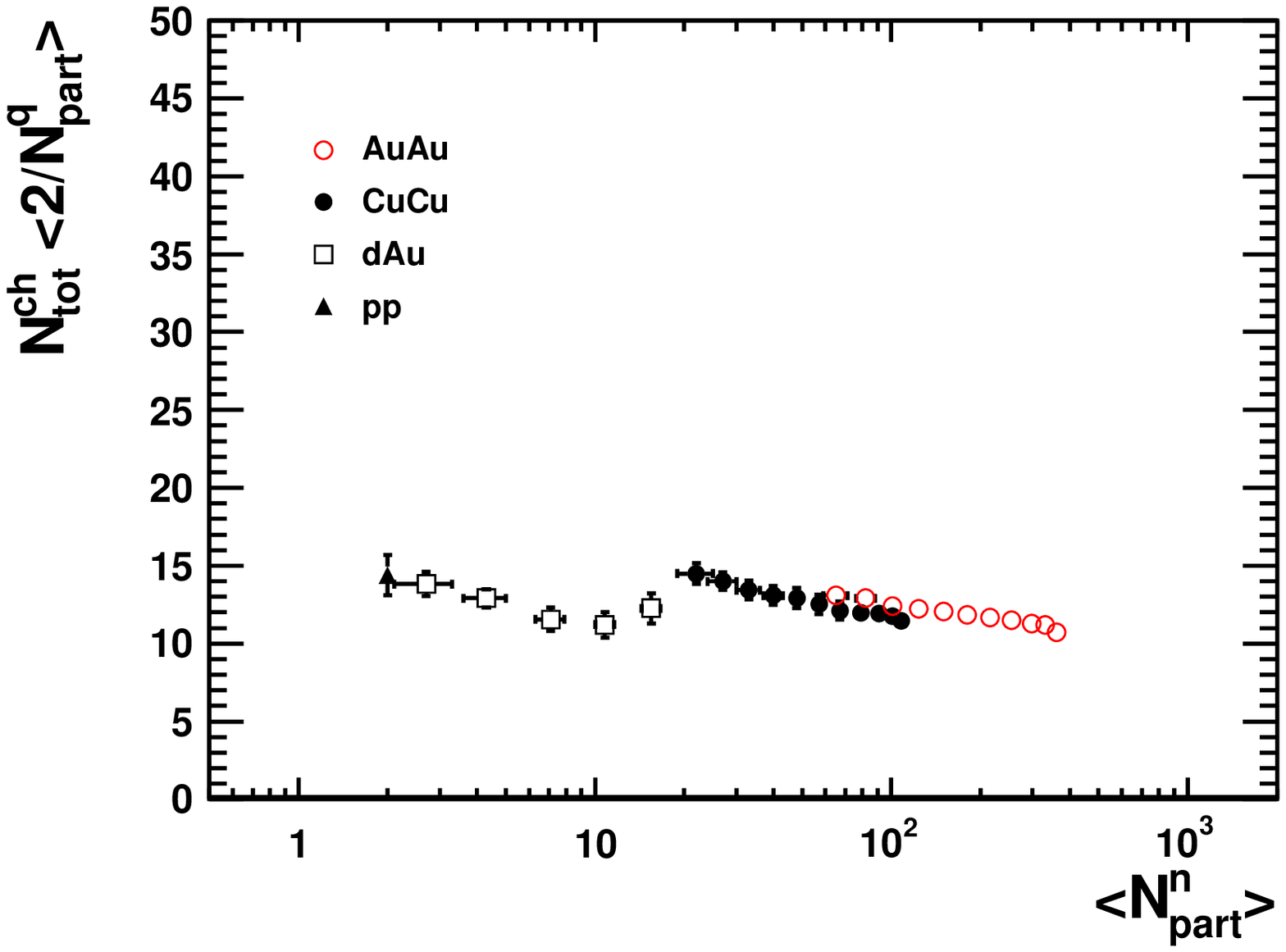}
\label{fig:Nch_tot_centrality_ncq}
}
\caption{(Color online) Total charged multiplicity varying with centrality in different collision systems scaled by the number of nucleon participants at $\sqrt{s_{NN}}=200$ GeV. 
Total charged multiplicity data for pp, dAu, CuCu and AuAu data are listed in~\cite{Alver:2010ck} . }
\label{fig:Nchtot_centrality}
\end{figure*}

The total multiplicity, $N^{\rm{ch}}_{\rm{tot}}$, for different collision systems at $\sqrt{s_{NN}}=200$ GeV is shown in Fig.~\ref{fig:Nchtot_centrality} by integrating the charged density measurement over the full rapidity range. The noticeable difference between the small collision systems including the pp and dAu values compared to AuAu values shown in Fig.~\ref{fig:Nch_tot_centrality_normal} is another argument against the nucleon participant picture. The ``leading particle effect" meaning the leading protons take away half the $pp$ center of mass energy missing from the multiplicity production has to be introduced to explain such a difference. 

Nevertheless, in the quark participant framework, the significant difference between pp/dAu collisions and CuCu/AuAu collisions has been partially removed as indicated by Fig.~\ref{fig:Nch_tot_centrality_ncq}. The total multiplicity scaled by quark participant numbers for different collision systems are to some extent around the same level. One can connect this consistency of multiplicity production to the underlying quark constituent picture. Also, this observation might shed some light on a different explanation to the "leading particle effect" when comparing the pp and $e^{+}e^{-}$ data~\cite{Back:2006yw}. In this framework, the missing energy taken away by the leading protons is incorporated by the spectator quarks which are not involved in the reaction.

\section{Impact of quark participants assumption on the initial states in heavy ion collisions}
\label{sec:impact}
The wounded nucleon model based on the Glauber Monte Carlo approach naturally incorporates the statistical spatial fluctuation of the collision geometry in heavy ion collisions on the event-by-event basis. Therefore, the wounded nucleon model and its extensions have been used as an important tool to simulate the initial states for the evolution in heavy ion collisions. It is then worthwhile to see the impact of replacing the nucleon participants with the quark participants on the initial states of heavy ion collisions.

\begin{figure}[htbp]
\centering
\includegraphics[width=0.55\textwidth]{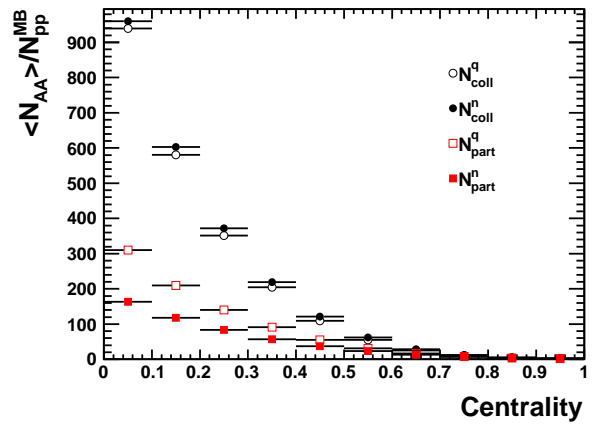}
\caption{(Color online) $\ncoll$ and $\npart$ in AuAu collisions normalized by that in minimum bias pp collisions for quark and nucleon participants. Open markers represent the result in quark constituent framework while the full markers are for nucleon constituent picture. }\label{fig:AA_Nq_Nn_divideby_pp}
\end{figure}

It is believed that the bulk and jet contribution are related to the number of participant $N_{part}$ and the number of binary collisions $N_{coll}$, respectively. The nuclear effect is usually extracted by studying the ratio of the same measurements from AA collisions and pp collisions after normalized by $N_{part}$ or $N_{coll}$ in each collision centrality class. We present the $N_{part}$ ($N_{coll}$) from AA collisions normalized by that from pp collisions in Fig.~\ref{fig:AA_Nq_Nn_divideby_pp}. One can observe that pp normalized $N_{part}$ in quark participant picture is around a factor of 2 of that in the wounded nucleon picture in the most central collisions.

\begin{figure*}[htbp]
\centering
\subfigure[]{
\includegraphics[width=0.45\textwidth]{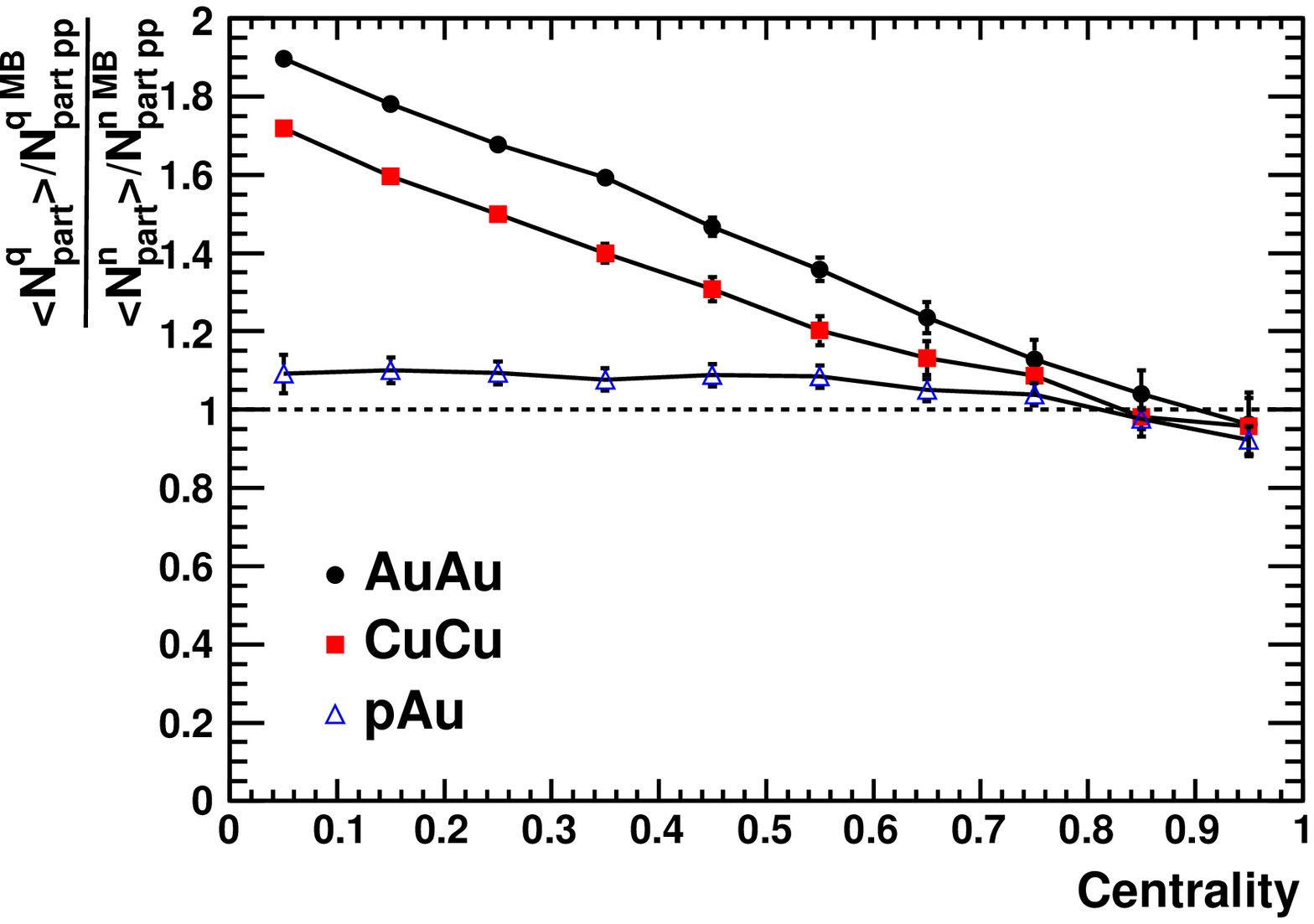}
\label{fig:npart_double_ratio}
}
\subfigure[]{
\includegraphics[width=0.45\textwidth]{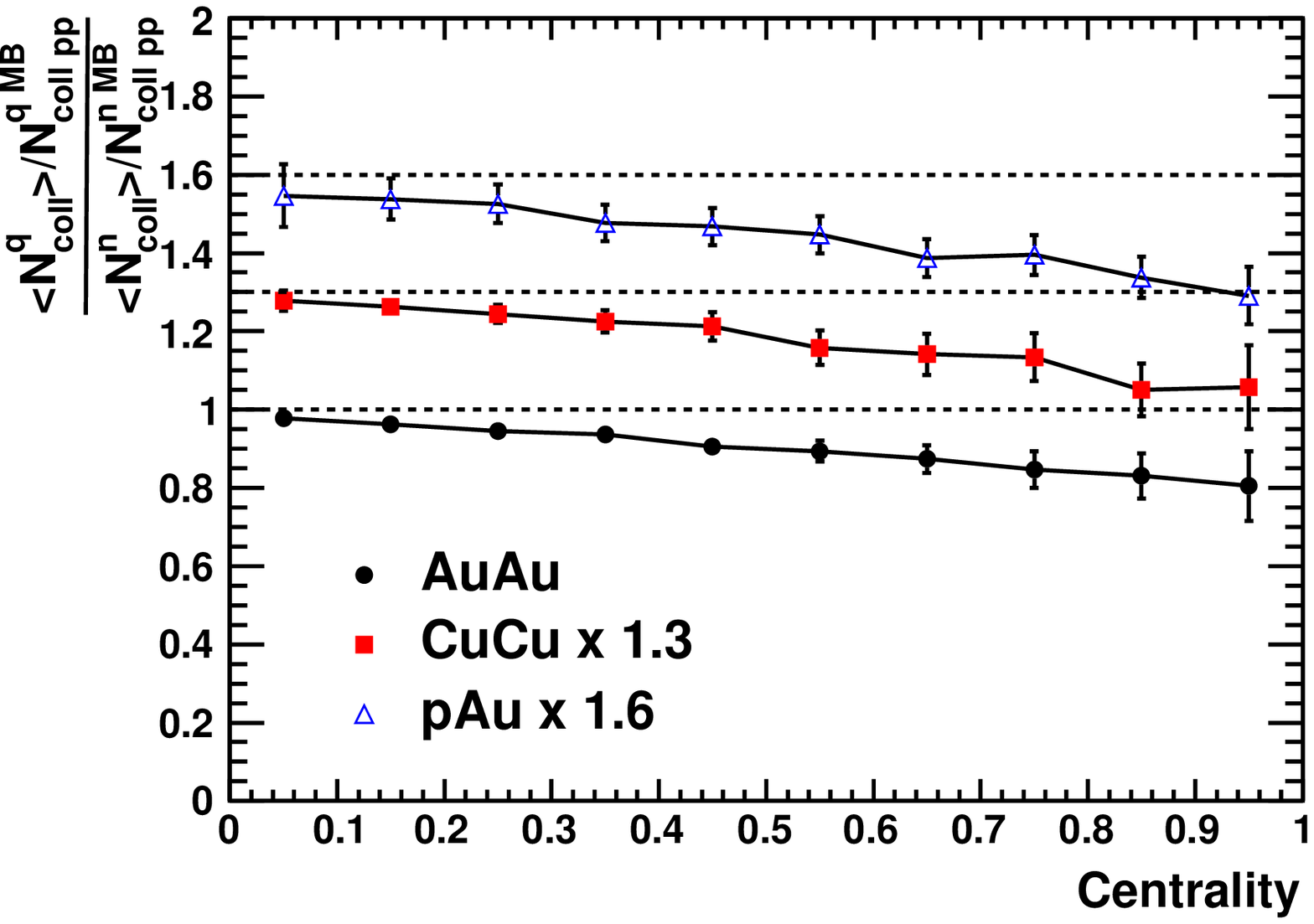}
\label{fig:ncoll_double_ratio}
}
\caption{(Color online) $\ncoll$ and $\npart$ double ratio varying with centrality in different collision systems at the energy of $\sqrt{s_{NN}}=200$ GeV. Normalization factor for pp were calculated in minimum bias events. AuAu results represented by solid dots, CuCu results shown in solid squares while pAu displayed in open triangles. $\ncoll$ for CuCu and pAu have been shifted by a factor of 1.3 and 1.6 for the purpose of demonstration.}\label{fig:double_ratio}
\end{figure*}

The centrality dependence of the impact from quark framework can be better visualized in Fig.~\ref{fig:double_ratio}, in which we make a double ratio of the pp normalized $\npart$ (Fig.~\ref{fig:npart_double_ratio}) and $\ncoll$ (Fig.~\ref{fig:ncoll_double_ratio}) between that from quark and nucleon participants.
If the double ratio value is around unity, we expect no modification in the related measurements when switching from the nucleon picture to quark picture. And this is exactly the case for $\npart$ in peripheral collisions and $\ncoll$ in central collisions.

A pronounced system size dependence can be observed in the $\npart$ distribution shown by Fig.~\ref{fig:npart_double_ratio}. From pAu, CuCu to AuAu collisions, we observe stronger enhancement for larger collision systems and all three curves converge to unity in very peripheral collisions. The impact from this feature has been shown in the multiplicity measurements in Sec.~\ref{sec:data_comparison}. On the other hand, little system size dependence can be found in the $\ncoll$ distribution shown in Fig.~\ref{fig:ncoll_double_ratio}. A 30\% deviation from unity can be found in the peripheral collisions for all systems, thereby introducing a small modification to the description of hard processes related measurements like nuclear modification factor. 

\begin{figure}[htbp]
\centering
\subfigure[]{
\includegraphics[width=0.45\textwidth]{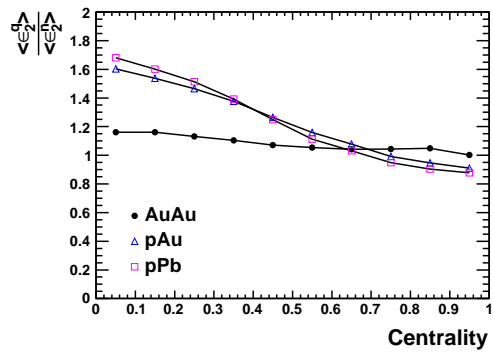}
\label{fig:Ecc_centrality}
}
\subfigure[]{
\includegraphics[width=0.45\textwidth]{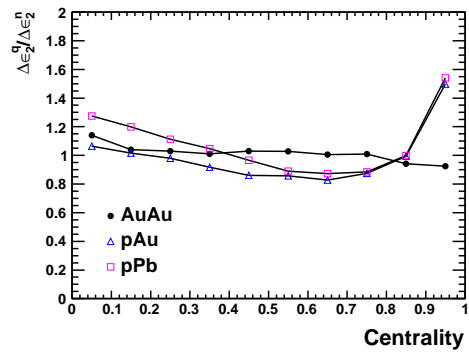}
\label{fig:fluc_Ecc_centrality}
}
\caption{(Color online) Participant eccentricity and its fluctuation based on quark participants divided by that based on nucleon participants varying with centrality in different collision systems. Calculations for AuAu (full dots) and pAu (open triangles) have been performed at the energy $\sqrt{s_{NN}}=200$ GeV,  and results for pPb (open squares) are obtained with $\sqrt{s_{NN}}=2.76$ TeV.}
\label{fig:Ecc2}
\end{figure}

It has been studied in~\cite{Rybczynski:2011wv,Heinz:2011mh} that the underlying nucleon profile is important for the initial geometry states used in hydrodynamic calculations. 
The quark participant picture used in our work can also be interpreted as a way to effectively model the sub-structure and fluctuation of the nucleon objects. Inspired by these discussions, we investigated the participant eccentricity defined as:
\begin{equation}
\epsilon = \frac{\sqrt{(<y^2>-<x^2>)^2+4<xy>^2}}{<y^2+x^2>},
\end{equation}
in which $x$ and $y$ determine coordinates of participating constituents in the transverse plane. The magnitude of this effect on the participant eccentricity and its fluctuation has been shown in Fig.~\ref{fig:Ecc2}, with quark participant eccentricity divided by nucleon participant eccentricity. We observe in Fig.~\ref{fig:Ecc_centrality} that $\epsilon$ in pPb and pAu collisions is quite sensitive to the introduction of sub-nucleon constituents, as the ratio shows around 60\% deviation from unity in the most central collisions, while AuAu exhibit slight variation to the quark participant mechanism. This feature can be understood as the nucleon-nucleon distance scale is much larger than the sub-nucleon length scale. The contribution of quark participants from the initial spatial geometry can only be observed when one of the colliding objects is a proton so that the initial collision geometry size is constrained to a level around the sub-nucleon scale. The effect of quark participant assumption on eccentricity fluctuation in Fig.~\ref{fig:fluc_Ecc_centrality} is found to be significant for small systems including pAu and pPb in the most peripheral collisions. This observation indicates that sub-nucleon spatial fluctuation plays an important role when the involved nucleon participant number is very small.

\section{Summary}
\label{sec:summary}
A systematic study of the geometric quantities from the quark participant based Glauber model has been presented in this work. The geometric scaling behavior of $\ncoll \propto \npart ^{4/3}$ can be observed in the quark participant model similar to that in the wounded nucleon assumptions which is required by the eikonal approximation in Glauber multiple scattering approach. It means that the quark participant formalism is a self-consistent approach to describe nuclear inelastic collisions. The multiplicity production has been examined in the quark participant model. A scaling like behavior of multiplicity production dependent on the number of quark participants can be found when confronted with the experimental data. 

We have also shown that the initial collision geometry is sensitive to the sub-nucleon structure with the application of quark participant approach. 
Initial state eccentricity of heavy ion collisions driving the magnitude of azimuthal flow coefficient from wounded quark assumption significantly deviates from that based on wounded nucleon assumption especially for small systems in central collisions. It is necessary to take this scale variation into regards when extracting the shear viscosity for pA collisions, as it is observed in~\cite{Noronha-Hostler:2015coa,Yan:2014nsa} that pPb collisions are found to be very sensitive to sub-nucleon scale fluctuations.

The idea of quark participants introduces a mechanism to describe the geometry fluctuations inside the nucleon, which may shed some light on the explanation of elliptic anisotropies recently observed in the high multiplicity, high energy pp collision events ~\cite{Aad:2015gqa}. The usage of quark participant assumption can be helpful in the analysis of small system when initial state fluctuations become important.

\begin{acknowledgement}
This work is supported by the NSFC (11475068) and the National Basic Research Program of China (2013CB837803). 
\end{acknowledgement}

\bibliography{reference}
\bibliographystyle{bibstyle/epjstyle}

\end{document}